\documentclass[preprint,showpacs,preprintnumbers,amssymb]{revtex4}

\usepackage{graphicx}
\usepackage{dcolumn}
\usepackage{bm}
\usepackage{feynmf}

\newcommand{\be}{\begin{equation}}
\newcommand{\ee}{\end{equation}}
\newcommand{\bea}{\begin{eqnarray}}
\newcommand{\eea}{\end{eqnarray}}

\newcommand{\GeV}{\rm GeV} 

\begin{document}

\bibliographystyle{apsrev}

\preprint{UAB-FT-522}

\title{On Axion Thermalization in the Early Universe} 

\author{Eduard Mass{\'o}} 
\email[]{masso@ifae.es}
\author{Francesc Rota}
\email[]{rota@ifae.es}
\author{Gabriel Zsembinszki}
\email[]{zgabi2001@yahoo.com}

\affiliation{Grup de F{\'\i}sica Te{\`o}rica and Institut 
de F{\'\i}sica d'Altes
Energies\\Universitat Aut{\`o}noma de Barcelona\\ 
08193 Bellaterra, Barcelona, Spain}



\begin{abstract}
We reanalyze the conditions under which we have a primordial thermal population of axions. We extend a previous study and take into account processes involving gluons and quarks. We conclude that if the Peccei-Quinn scale fulfills $F_a < 1.2 \times 10^{12}\,\GeV$ there is thermal axion production. In this case, a period in the early universe exists where axions would interact with the QCD plasma and we point out that non-thermal axions produced before the end of this period would thermalize. 
\end{abstract}

\maketitle


\section{Introduction}
\label{intro}

The Peccei-Quinn solution \cite{PQ} to the strong CP-problem
introduces a new $U(1)_{PQ}$ chiral symmetry, the spontaneous breaking of which
leads to a new spinless particle, the axion \cite{WW}. 
If it exists, the axion has quite well defined properties: its
mass and couplings are inversely proportional to $F_a$, the energy
scale of the $U(1)_{PQ}$ breaking. We will be particularly concerned with
the coupling of the axion field $a$ to gluons
\be
{\cal L }_{gga} = \frac{1}{F_a}\, \frac{\alpha_s}{8\pi} \,
G^{b\mu\nu}\, \widetilde G^b_{\mu\nu}\ a
\label{gga}
\ee
where $\alpha_s=g_s^2/(4\pi)$ with $g_s$ the gauge coupling of 
color $SU(3)_c$. In (\ref{gga}),
\be
\widetilde G^b_{\mu\nu}= \frac{1}{2}\, \epsilon_{\mu\nu\rho\sigma}
G^{b\rho\sigma}
\label{dual}
\ee
is the dual of the gluon field $G^b_{\mu\nu}$, and 
there is a sum over the color index $b$.

Much theoretical study and experimental effort have been devoted to this particle (for recent reviews see \cite{review}). The laboratory, astrophysical, and cosmological constraints imply that $F_a$ must be a high energy scale, at least as large as \cite{Raffelt}
\be
 F_a > 6 \times 10^8\, \GeV
\label{boundSN}
\ee
which in turn implies
\be 
m_a < 0.01\ {\rm eV}
\label{mass}
\ee
Thus, the axion has to be a very light particle and very weakly coupled.

One of the most attractive features of the axion is that
it may be one of the constituents of the dark matter.
Indeed, axions are copiously produced in the early universe. These
primordial axions are generated mainly non-thermally. 
We know three non-thermal sources 
at work in the evolution of the universe: the ``misalignment mechanism'' 
\cite{misalignment}, string decay \cite{string} and domain wall decay \cite{Chang}. When the Peccei-Quinn symmetry breaks down at a
temperature $T\simeq F_a$, the vacuum angle has no preferred value;
but when the temperature reaches $T\simeq \Lambda_{QCD}$, the
vacuum angle sets where the vacuum energy is minimized, which is the
CP-conserving value. Relaxation from one angle to the other 
produces a coherent, cold condensate of axions. This is the  
``misalignment mechanism''. On the other hand, the $U(1)_{PQ}$ breaking
at $T \simeq F_a$ originates axion strings. If inflation
takes place just after that, these strings dilute away. However,
if this is not the case, we have a second mechanism of production
of axions when strings decay. Also, if the axion field is not homogeneized by inflation, there are models where axion domain walls appear soon after $T \simeq \Lambda_{QCD}$. The decay of these walls into axions is a third mechanism of non-thermal primordial axions.

Thermal production of axions was studied by Turner \cite{Turner}.
His main interest was to study the possibility that thermal
axions have a density greater than the corresponding to non-thermal
axions, and then to analyze the potential detection of photons from
axion decay.
Turner considered the Primakoff process, $a + q \rightleftarrows \gamma + q$, 
where $q$ is a light quark, and photoproduction,
 $a + Q \rightleftarrows \gamma + Q$, where $Q$ is a heavy quark,
and showed that when
\be
 F_a <  10^9\, \GeV
\label{bound_turner1}
\ee
axions were once in thermal equilibrium. Evaluating the axion
density he found that for
\be
 F_a < 4 \times 10^8\, \GeV
\label{bound_turner2}
\ee
the thermal population is greater than the non-thermal.

In \cite{Turner}, it was mentioned that other axion processes contribute to axion thermalization
and consequently would increase the value in (\ref{bound_turner1}).
However, these other processes were not evaluated since the range 
(\ref{bound_turner1}) includes (\ref{bound_turner2}) and this was enough
for the purposes of \cite{Turner}. 
We note that the experimental bound in (\ref{boundSN}) excludes the
range in (\ref{bound_turner2}) and leaves a small interval of
the condition (\ref{bound_turner1}).

In this paper we extend the work of \cite{Turner} 
and investigate the contribution to axion thermalization 
of other processes than Primakoff and photoproduction. As we will show, the
processes we discuss lead to a much larger range for $F_a$ than
(\ref{bound_turner1}). 

The reactions we consider all involve the $gga$ vertex 
appearing in (\ref{gga}),

1) $a + g  \rightleftarrows q + \bar q$

2) $a + q \rightleftarrows g + q $ \ 
and \ $a + \bar q \rightleftarrows g + \bar q $

3) $a + g \rightleftarrows g + g$

\begin{figure}
\begin{center}
\includegraphics{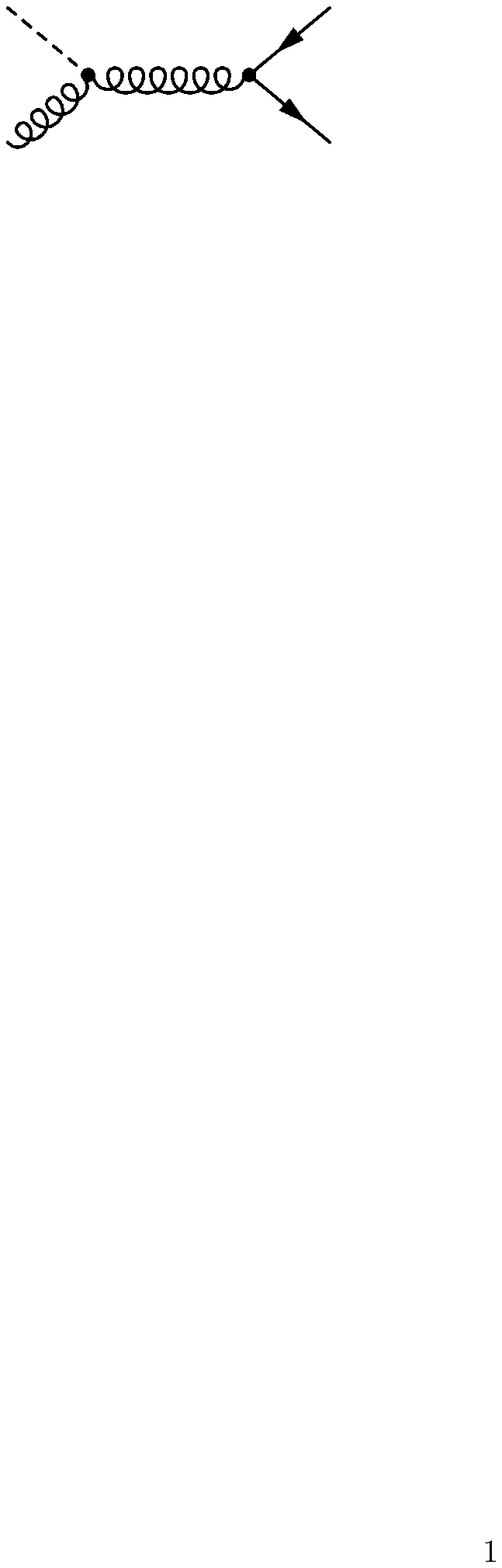}
\end{center}
\caption{\label{fig_1} Feynman diagram for the process $a + g \rightleftarrows  q + \bar q$.} 
\end{figure}

\begin{figure}
\begin{center}
\includegraphics{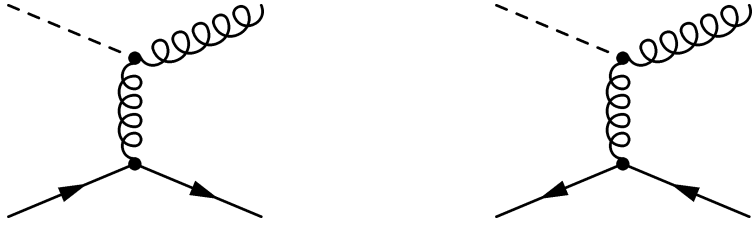}
\end{center}
\caption{\label{fig_2} Feynman diagrams for the processes $a + q \rightleftarrows g + q$ and $a + \bar q \rightleftarrows g + \bar q$.} 
\end{figure}

\begin{figure}
\begin{center}
\includegraphics{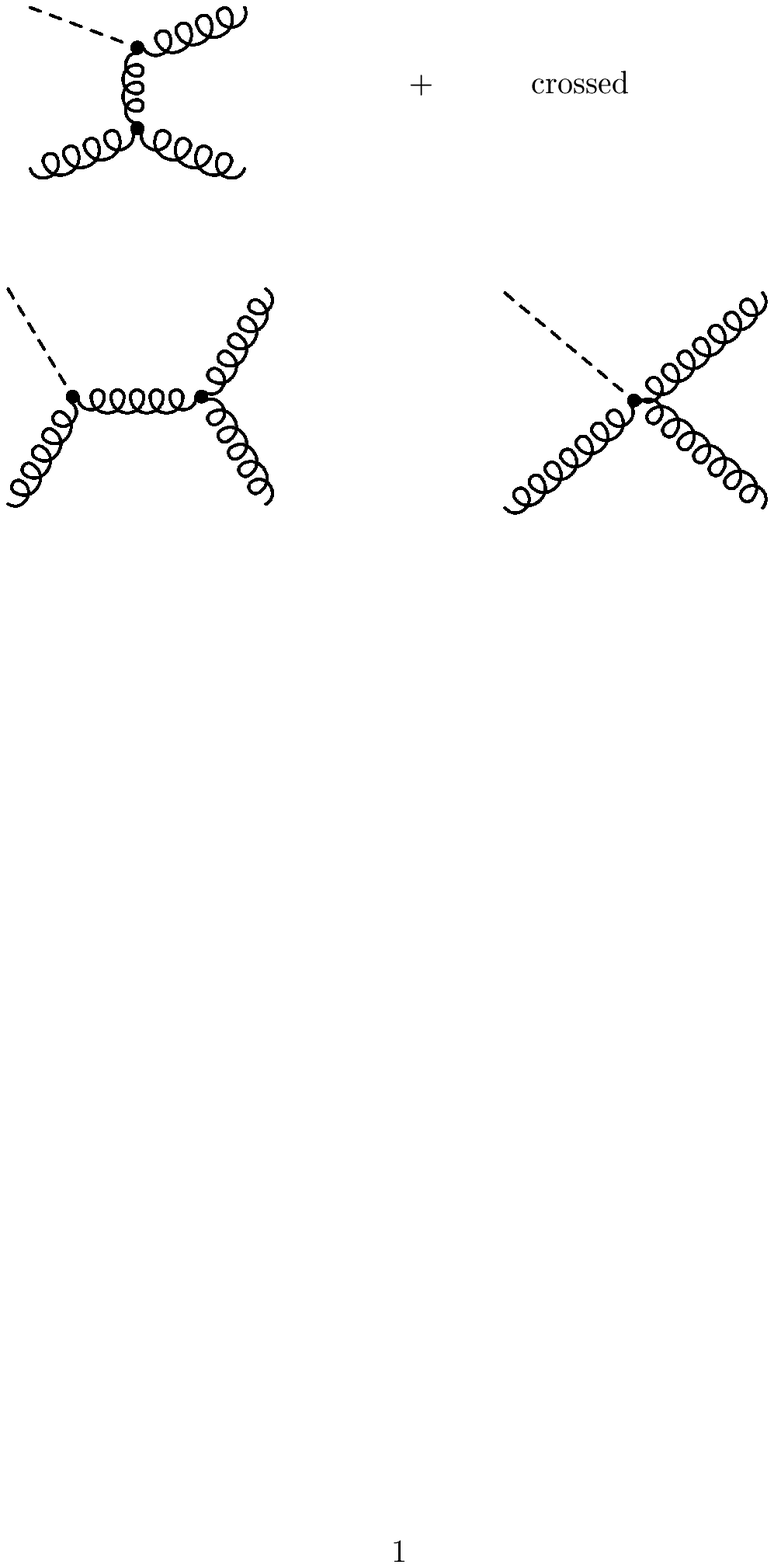}
\end{center}
\caption{\label{fig_3} Feynman diagrams for the process  $a + g \rightleftarrows g + g$.} 
\end{figure}

We show the corresponding Feynman diagrams in 
Figs. \ref{fig_1}, \ref{fig_2}, and \ref{fig_3}.

The coupling of the axion to gluons (\ref{gga}) is an essential ingredient
of the effective theory once the Peccei-Quinn symmetry is broken,
since it reproduces the chiral anomaly of the theory. In this sense
our results are model independent in contrast with the result
(\ref{bound_turner1}) arising from Primakoff and photoproduction processes 
that involve the $a\gamma\gamma$ coupling
which is model dependent and might be small in some models.

\section{Evolution of the axion density}
\label{axion_density}

In the cooling of the universe, as soon as the temperature $T \simeq F_a$
is reached, massless axions can be produced by means of
reactions 1), 2), and 3). The production rate has to compete
with the expansion of the universe, characterized by
the Hubble expansion rate,
\be
H = \sqrt{ \frac{4\pi^3}{45} }\, \sqrt{g_*}\, 
\frac{T^2}{M_{planck}}
\label{hubble}
\ee

We will assume that, in the range of temperatures where we will apply the evolution equation for interating axions, the particle content corresponds to $SU(3)_c \times SU(2)_L \times U(1)_Y$, with 3 families
and one Higgs boson, with all particles at the same temperature. 
Then, the relativistic degrees of freedom appearing in (\ref{hubble})
are given by \cite{KolbTurner}
\be
g_*= g_{boson} + \frac{7}{8}\, g_{fermion} = 106.75
\label{g}
\ee

The behaviour of the axion density $n_a$ is given in terms of
the Boltzmann equation \cite{Turner,KolbTurner,Brown},
\be
\frac{d\, n_a}{dt} + 3 H n_a = \Gamma\, [n_a^{eq} - n_a]
\label{bol1}
\ee
Here,  $\Gamma$ is the thermal averaged interaction rate for the process $a + i \rightleftarrows 1 + 2$
\bea
\Gamma &\equiv& \frac{1}{n_a^{eq}}\int \,\frac{d\tilde p_a}{2E_a}\,\frac{d\tilde p_i}{2E_i}\,f_a^{eq}\,f_i^{eq}\cdot (\tilde \sigma\, v\, 2E_a2E_i)
\label{gamma}
\\
\tilde \sigma\, v\, 2E_a2E_i &\equiv& \int \,\frac{d\tilde p_1}{2E_1}\,\frac{d\tilde p_2}{2E_2}\,(2\pi)^4\,\delta^4(p_1+p_2-p_a-p_i)\,\sum \left|M\right|^2
\label{sigma}
\eea
with $d\tilde p \equiv d^3p/(2\pi)^3$. The phase space occupancy in kinetic equilibrium $f^{eq}$ is given by the Fermi-Dirac or Bose-Einstein distributions,
\be
f^{eq} = \frac{1}{e^{(E-\mu)/T}\pm 1}
\label{distribution}
\ee
where we will take the relativistic limit where the chemical potential $\mu$ goes to zero. In eq (\ref{sigma}), $\left|M\right|^2$ is summed over all the degrees of freedom. The equilibrium densities appearing in (\ref{gamma}) are given by
\be
n^{eq} = g \int d\tilde p \,f^{eq}
\ee
where $g$ is the degrees of freedom of the particle. In the period of interest, thermal axions are massless and have a density
\be
n_a^{eq} = \frac{\zeta(3)}{\pi^2}\,T^3
\label{eq}
\ee
where $\zeta(3) = 1.20206\ldots$ is the Riemann zeta function of 3. 

We follow the convenient procedure 
\cite{Turner,KolbTurner,Brown} of normalizing particle densities
to the entropy density,
\be
s = \frac{2 \pi^2}{45} \, g_{*S}\, T^3 = 0.44\times g_{*S}\, T^3
\label{entropy}
\ee
so that we define
\be
Y \equiv \frac{n_a}{s}  
\ee
During the epoch in which we will apply the Boltzmann equation (\ref{bol1}), we can use $g_{*S}=g_*=$ constant, and given by (\ref{g}). Then we have that
\be
Y^{eq}=\frac{n_a^{eq}}{s}=
\frac{45\, \zeta(3)}{2\pi^4}\,\frac{1}{g_*} = \frac{0.27}{g_*}
\label{Yeq}
\ee
is constant. We can write the Boltzmann equation (\ref{bol1})
as
\be
x\, \frac{d\, Y}{dx} = \frac{\Gamma}{H}\, 
 \left[ Y^{eq} - Y \right]
\label{bol2}
\ee
where we have defined
\be
x \equiv \frac{F_a}{T}
\ee 

We will now show that equation (\ref{bol2}) can be integrated in our
case.
First, we advance that $\Gamma\propto T^3$ (see sec. \ref{calcul}) so that 
$\Gamma/H \propto 1/x$ and we can define the constant
\be
k \equiv x\, \frac{\Gamma}{H}
\label{k}
\ee
Second, we use the fact that
$Y^{eq}$ is independent of $x$.
We define 
\be
\eta \equiv \frac{Y}{Y^{eq}}
\ee
and, finally, we write equation (\ref{bol2}) in the form
\be
x^2 \frac{d\, \eta}{dx} = k\,  (1-\eta)
\label{bol3}
\ee
that has the solution
\be
\eta(x) = 1-e^{k(1/x -1)}
\label{sol}
\ee
Since we start at $x=1$ ($T=F_a$) with no axions, in (\ref{sol})
we have specified  the initial condition $\eta(x=1)=0$.

From the solution (\ref{sol}) we see that $\eta$ starts to grow at $x=1$. At $x=k$ (that is, $\Gamma=H$) axions decouple from the QCD plasma. We call $Y_d$ the value of $Y$ at decoupling. For $x>k$ the value of $\eta$ remains constant, so we have $Y=Y_d$. We are interested in the situation that axions have in practice a thermal spectrum, so we will ask that $Y_d$ differs from $Y^{eq}$ in less than 5\%,
\be
\frac{Y_d}{Y^{eq}} = \eta(x=k) = 1-e^{k(1/k -1)} > 0.95
\ee   
which implies
\be
k> 4
\label{condition}
\ee
When $k$ satisfies this inequality, a thermal population of axions is
born in the early universe. In the next section we calculate 
$\Gamma$, which, as we see from (\ref{k}), will give us the clue
of which is the corresponding range of $F_a$.

\section{Calculation of thermally averaged interaction rates} 
\label{calcul}

As expression (\ref{sigma}) is Lorentz invariant, we can evaluate it in any reference system. For convenience, we choose the center of mass system and will consider that all the particles are massless because we are at very high energies. So, we will write (\ref{gamma}) as
\be
\Gamma = \frac{1}{n_a^{eq}}\int \,\frac{d\tilde p_a}{2E_a}\,\frac{d\tilde p_i}{2E_i}\,f_a^{eq}\,f_i^{eq}\cdot(2\,s\,\tilde\sigma_{CM})
\label{gammaCM}
\ee
where $\tilde\sigma_{CM}$ is the usual total cross section in the center of mass system, but with no average over the initial degrees of freedom, and $s$ is the Mandelstam invariant. From the Lagrangian piece (\ref{gga}) one can easily find the Feynman rules for the couplings between axions and gluons that appear in the diagrams in Figs.\ref{fig_1}, \ref{fig_2}, and \ref{fig_3}, which are necessary to calculate the cross sections for the considered processes. Some of these cross sections diverge logarithmically in the $t$ and $u$ channel. The processes we consider take place in a plasma with vanishing global color and so color is effectively screened for distances bigger than $m_D^{-1}$, with $m_D$ the QCD Debye mass given by \cite{bellac}
\be
m_D^2 = g_s^2\, \frac{N_c+N_f/2}{3}\, T^2 = 8\pi\alpha_s\,T^2
\ee 
where $N_c = 3$ is the number of colors and $N_f = 6$ is the number of flavors. We cutt-off the divergences using the Debye mass. Cross sections have the form
\be
\tilde\sigma_{CM} = A \ln{\left(\frac{s}{m_D^2}\right)} + B
\label{sigmaCM}
\ee 
where $A$ and $B$ are constants, which for the considered processes have the numerical value

1) $ ~~ A = 0 ~~~~~,~~~~~ B = \frac{N_f}{6\pi^2}\,\frac{\alpha_s^3}{F_a^2}$

2) $ ~~ A = \frac{N_f}{\pi^2}\,\frac{\alpha_s^3}{F_a^2} ~~, ~~ B = -\frac{3\,N_f}{4\pi^2}\,\frac{\alpha_s^3}{F_a^2}$

3) $ ~~ A = \frac{15}{2\pi^2}\,\frac{\alpha_s^3}{F_a^2} ~~, ~~ B = -\frac{55}{8\pi^2}\,\frac{\alpha_s^3}{F_a^2}$\\
Putting (\ref{sigmaCM}) in (\ref{gammaCM}) we find integrals of the type
\bea
I^{\pm} &\equiv& \frac{1}{\zeta(3)}\,\int_0^{\infty}dx\,\frac{x^2}{e^x \pm 1} \\
L^{\pm} &\equiv& \frac{1}{\zeta(3)}\,\int_0^{\infty}dx\,\frac{x^2\,\ln x}{e^x \pm 1}
\eea
that can be done analitically. We have
\bea
I^+ &=& \frac{3}{2}\\
I^- &=& 2 \\
L^+ &=& \frac{1}{4} \left(9+\ln 4 - 6\gamma + 6\frac{\zeta'(3)}{\zeta(3)} \right)\\
L^- &=& 3 - 2\gamma + 2\frac{\zeta'(3)}{\zeta(3)} 
\eea
where $\gamma = 0.5772\ldots$ is the Euler's Gamma constant and $\zeta'(3) = -0.1981\ldots$. With these results, the integral (\ref{gammaCM}) gives
\be
\Gamma = \frac{T^3\,\zeta(3)}{(2\pi)^2} \left\{ \left[B-A\left(1/2+\ln(2\pi\alpha_s)\right)\right]\,I^{\pm}\,I^- + A\,L^{\pm}\,I^- + A\,I^{\pm}\,L^- \right\}
\ee
where the sign $\pm$ depends on whether the particle that goes whith the axion is a gluon ($-$) or a quark ($+$). Thus, for each of the processes 1), 2), and 3) we get, respectively,
\bea
\Gamma_1 &\equiv& \Gamma\,(a + g \rightleftarrows q + \bar q) = \frac{\alpha_s^3}{F_a^2}\,T^3\,N_f\,\frac{\zeta(3)}{6\pi^4}\\
\Gamma_2 &\equiv& \Gamma\,(a + q \rightleftarrows g + q) + \Gamma\,(a + \bar q \rightleftarrows g + \bar q) = \nonumber \\
 && \frac{\alpha_s^3}{F_a^2}\,T^3\,2N_f\,\frac{\zeta(3)}{4\pi^4}\,\left(L^++\frac{3}{4}L^--\frac{3}{2}\ln{(2\pi\alpha_s)}-\frac{15}{8} \right)\\
\Gamma_3 &\equiv& \Gamma\,(a + g \rightleftarrows g +g) = \frac{\alpha_s^3}{F_a^2}\,T^3\,\frac{15\,\zeta(3)}{2\pi^4}\,\left(L^--\ln{(2\pi\alpha_s)}-\frac{17}{12} \right)
\eea
$\Gamma$ is the sum of all $\Gamma_i$, and turns out to be
\be
\Gamma \simeq 7.1 \times 10^{-6}\,\frac{T^3}{F_a^2}
\label{GammaT}
\ee
where we have introduced $\alpha_s \simeq 1/35$, corresponding to energies $E \simeq 10^{12} \,\GeV$. 

\section{Conclusions}
\label{concl}
Using (\ref{GammaT}), (\ref{hubble}) and the definition (\ref{k}) we have
\be
k = \frac{F_a}{T}\,\frac{\Gamma}{H} \simeq 5.0\times 10^{12}\,\frac{\GeV}{F_a}
\ee
The inequality (\ref{condition}) translates into
\be
F_a < 1.2 \times 10^{12} \,\GeV
\label{result}
\ee
This is our main result. Models with $F_a$ satisfying (\ref{result}) have axions thermalizing in the early universe and predict that today there is a thermal population of axions.

We get a much higher value
than the one in (\ref{bound_turner1}). One of the reasons is that
the interaction rate (\ref{GammaT}) increases as $T^3$ at high energies,
while the rate coming from the photoproduction process considered 
in \cite{Turner}
goes as $T$ since the axion is attached to the fermion line with the usual derivative coupling. The other reason is that
color and flavor factors make the interaction rate larger.

Our result in (\ref{result}) is in fact conservative. 
This is due to the fact that the effective theory below the scale $F_a$ has couplings of the axion to the electroweak gauge bosons $W^{\pm}$, $W^0$, and $B$, similar to (\ref{gga}). This generates processes like 1), 2), and 3) with gluons replaced by appropiate combinations of $W^{\pm}$, $W^0$, and $B$. These new processes contain in general quarks and also leptons. The form of the couplings is fixed but their magnitude is model dependent since it is different for different PQ-charge assignments to the matter fields. Given this model dependence, we do not include them in our analysis. They would lead to an increase of the numerical value in (\ref{GammaT})  by a factor of $\sim 2$ and consequently would make the range of $F_a$ in (\ref{result}) larger by the same amount.

Once we assume that axions thermalize, it is interesting to know the density of this thermal axion population. After decoupling, axions redshift freely until today. Their density today, $n_{a0}$, can be calculated using the fact that after decoupling, the value of $Y$ stays constant. We need the entropy density today, $s_0$,
which has the contribution of photons, neutrinos and axions. Even in the case that axions are relativistic today, the contribution to $s_0$ can be neglected since the axion temperature is much colder than the cosmic microwave background temperature, $T_0 = 2.75$ K. Indeed, since the time of axion decoupling, the decoupling of other particles have heated the photon bath but not the axion one. One gets that the entropy density today 
corresponds to $g_{*S}= 43/11 = 3.91$ in (\ref{entropy}).

Then, we use (\ref{g}), (\ref{entropy}), and (\ref{Yeq}), to calculate the axion number density today
\be
n_{a0} = \frac{0.27}{106.75}\,  (0.44\times 3.91\, T_0^3) \simeq 7.5 \, \rm cm^{-3}
\label{n_today}  
\ee
This small density was also found by Turner \cite{Turner}. At the view of the small mass (\ref{mass}) one can conclude that it is very unlikely that it can play a direct role in the efforts to detect axions \cite{review}. 

It is also interesting to determine the range of temperatures where axions interact with the QCD plasma. For a given $F_a$ satisfying (\ref{result}), it is not difficult to see that the thermalization range is 
\be
F_a~ \gtrsim~ T ~\gtrsim ~\frac{F_a^2}{5\times 10^{12}\,\GeV}
\label{range}
\ee
For example, for the highest possible value of $F_a$ where still a thermal population arises, $F_a = 1.2 \times 10^{12}\,\GeV$, we get a temperature range
\be
1.2 \times 10^{12} \,\GeV ~\gtrsim~ T ~\gtrsim~0.3\times 10^{12}\,\GeV
\ee
For the lowest value of $F_a$ allowed by observation (see eq(\ref{boundSN})), the range is
\be
6 \times 10^{8} \,\GeV ~\gtrsim ~ T ~\gtrsim~ 7\times 10^{4}\,\GeV
\label{range2}
\ee

Our final comment refers to a related effect also happening in the thermalization  period.
If there are non-thermal axions produced at this period,
they would thermalize since they would interact with the QCD plasma
through reactions 1), 2) and 3).
Thus, no matter which was the energy spectrum and how high was the production rate of the originally non-thermal axions, they would end up with the thermal spectrum (\ref{distribution}) and density (\ref{eq}). There is no effect on axions from the misalignment mechanism and domain wall decay, since 
they are generated at $T\simeq \Lambda_{QCD}$, that cannot be in the 
range (\ref{range}).
Strings start to decay into axions after their formation at $T\simeq F_a$.
Current models \cite{string}
have the bulk of axions produced also at $T\simeq \Lambda_{QCD}$.
If this is the case, then thermalization effects on string axions are small.
However, the physics of axion strings is still not completely settled and we would like
to stress that if a string model or any other source
of non-thermal axions have substantial axion production in the range
(\ref{range}) then those axions would thermalize and finally we would
have a single thermal population that dilute away with the expansion
of the universe as we have studied in this paper.

\begin{acknowledgments}
We thank Cristina Manuel and Georg Raffelt for discussions. The work of E. M. and F. R. is partially supported by the CICYT Research Project AEN99-0766, by the DGR Project 2001 SGR 00188, and by the EU network on {\it Supersymmetry and
the Early Universe} (HPRN-CT-2000-00152). G. Z. is grateful for
hospitality at the Grup de Fisica Teorica of the Universitat
Autonoma de Barcelona.
\end{acknowledgments}




\newpage

\end{document}